\documentclass[aps,pra,twocolumn,showpacs,preprintnumbers,superscriptaddress,amsmath]{revtex4}
\usepackage{graphicx}
\usepackage{dcolumn}
\usepackage{bm}

\begin{document}

\title{Engineering of multi-dimensional entangled states of photon pairs using
hyper-entanglement}
\author{Xi-Feng Ren, Guo-Ping Guo \footnote[2]{harryguo@mail.ustc.edu.cn}, Jian
Li, Chuan-Feng Li and Guang-Can Guo}
\address{Key Laboratory of Quantum Information, University of Science and\\
Technology of China, CAS, Hefei 230026, People's Republic of
China\bigskip }
\begin{abstract}
Multi-dimensional entangled states have been proven to be more
powerful in some quantum information process. In this paper,
down-converted photons from spontaneous parametric down
conversion(SPDC) are used to engineer multi-dimensional entangled
states. A kind of multi-degree multi-dimensional
Greenberger-Horne-Zeilinger(GHZ) state can also be generated. The
hyper-entangled photons, which are entangled in energy-time,
polarization and orbital angular momentum (OAM), is proved to be
useful to increase the dimension of systems and investigate
higher-dimensional entangled states.

\end{abstract}

\maketitle

\section {introduction}
Quantum entanglement is the foundation of quantum teleportation,
quantum computation, quantum cryptography, superdense coding, etc.
In recent years, the interest in multi-dimensional entangled
states, or qudits, is steadily growing. One advantage of using
multilevel systems is its promise to realize new types of quantum
communication protocols\cite{Bartlett00,Bechmann00,Bourennane01},
and its better properties in quantum cryptography than
qubits\cite{Bechmann00,Bourennane01,BechmannA00,Guo02}. These
cryptography protocols are more robust against specific classes of
eavesdropping attacks. The other advantage is their possible
implementation in the fundamental tests of quantum mechanics. For
two-partite system of dimension greater than two, it has been
found that the Clauser-Horne-Shimony-Holt(CHSH) inequality can be
maximally violated, and this violation continues to survive in the
limit of infinite dimensions\cite{Gisin92}. GHZ paradoxes and Bell
inequality have also been discussed for multi-partite systems of
multi-dimension in\cite{Zukowski99,Cerf02,Kaszli02,Acin04,Jinh04}.
Additionally, the usage of multilevel systems provides a
possibility to introduce very special protocol, which cannot be
implemented with qubits, such as quantum bit
commitment\cite{Langford04}.
\begin{figure}[b]
\includegraphics[width=8cm]{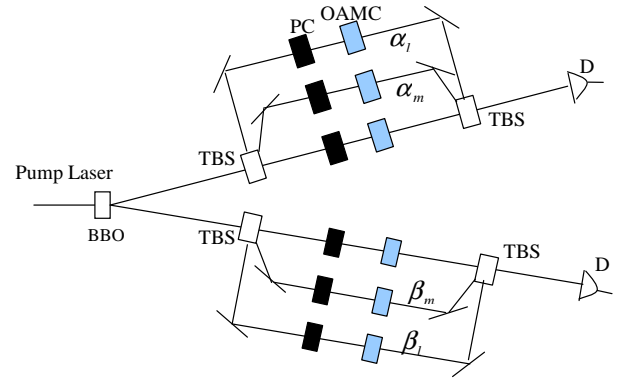}
\caption{ Schematic experimental setup for generation of
multi-freedom multi-dimensional entangled states. In each path,
there is a three-arm interferometers. TBS is three-beam-splitter
which can also be substituted by two BS. PCs and OAMCs are added
in each arm to control the polarization and OAM of down-converted
photons.}
\end{figure}

There are several approaches to investigate higher dimensional
systems. In one case, multiple entangled qubit systems are used,
such as four-photon polarization scheme\cite{Howell02}. In some
other cases, dimensions of the element are increased. For example,
the interferometer method was used to generate energy-time
entangled qutrits\cite{Thew04,Thew042}; and other techniques rely
on the spatial modes of down-converted photons from SPDC. These
down-converted photons can be entangled in not only polarization,
or spin angular momentum, but also spatial modes, such as orbital
angular momentum(OAM)\cite{Mair01}, Hermite-Gaussian
modes\cite{Ren042}. The spatial entanglement occurs in an
infinite-dimensional Hilbert space. Many
theoretical\cite{Arnaut00,Molina02,Torres04,Ren04} and
experimental\cite {Mair01,Vaziri02,Vaziri03} works about entangled
qudits have been done based on OAM of the photons.

Quantum state engineering, i.e., the ability to generate, transmit
and measure quantum systems is of great importance in quantum
information process. Qutrit state engineering has been done using
polarization of biphotons\cite{Boganov04}. In the present
protocol, we show that any entangled qudit states can be generated
by exploiting the OAM and polarization of the energy-time
entangled photons. A kind of artificial multi-partite
multi-dimensional entangled GHZ states is also generated by the
help of photon pairs entangled in polarization, OAM and
energy-time simultaneously.(This kind of multiply-entanglement is
named as hyper-entanglement\cite{Kwiat97}.) Although only two
photons are involved in these GHZ states, they appear like
multi-partite entangled states due to the usage of multi-degree of
each photon. We call this kind of states multi-degree
multi-dimensional entangled states.

\section {experimental setup to generate entangled qudits and multi-degree multi-dimensional GHZ states}

Energy-time entangled qutrit states can be generated as photon
pairs from SPDC passing two three-arm
interferometers\cite{Thew04,Thew042}. For each interferometer, a
phase vector consisting of two independent phases can be defined,
e.g., the relative phase between the short(s)-medium(m) and
short-long(l), path lengths. Coincidence measurement at the
outputs of the interferometers project onto entangled qutrit
states defined when the photons take the same arm in each
interferometer, short-short, medium-medium or long-long at
signal-idler path. For this type of experiments, the following
conditions must be satisfied\cite{Thew042}: The coherence length
of the down-converted photons is much smaller than the path-length
difference in the interferometers so that no single photon
interference effects are observed in passing the interferometers;
the coherence length of the pump laser is much greater than these
path-length differences so that we have no timing information as
to the creation time of the photon pairs and hence which path was
taken before detection.

As polarization control(PC) and OAM control (OAMC) are added, we
extents the experimental set-up of Thew and his
co-workers\cite{Thew04,Thew042} to Fig. 1. Consider the case that
the photons take the same path in each interferometer, the state
can be written as:

\begin{eqnarray}
\left| \psi \right\rangle \propto &&c_s\left| ss\right\rangle
+c_me^{i(\alpha _m+\beta _m+\Phi _{jk}^m)}\left| mm\right\rangle  \nonumber
\\
&&+c_le^{i(\alpha _l+\beta _l+\Phi _{jk}^l)}\left| ll\right\rangle ,
\end{eqnarray}
where $\alpha _m,\alpha _l$ and $\beta _m,\beta _l$ represent the
phase in medium and long interferometer arms of signal and idler
path, $\Phi _{jk}^m$ and $\Phi _{jk}^l$ are multiples of $2\pi /3$
which depend on the path taken by the photons in the
interferometer and the output, $j,k\in \{0,1,2\}_{A,B}$, they
taken. Apparently, we can justify the state of Eq. (1) to:

\begin{eqnarray}
\left| \psi \right\rangle &=&\frac 1{\sqrt{3}}(\left| ss\right\rangle
+\left| mm\right\rangle +\left| ll\right\rangle )  \nonumber \\
&=&\frac 1{\sqrt{3}}(\left| 00\right\rangle +\left| 11\right\rangle +\left|
22\right\rangle ).
\end{eqnarray}
It is a maximally entangled GHZ state of two
qutrits\cite{Thew04,Thew042}.

\subsection {Generation of qudits and three-degree three-dimensional GHZ state }
The Polarization control(PC)(see Fig. 2a) is a device that can be
used to control the polarization of photons. This can be realized
by the combination of Polarization Beam Splitter(PBS), Half Wave
Plate(HWP) and Quarter Wave Plate(QWP). The polarization of
photons passing through PBS can be determinately H or V. Using HWP
and QWP, we can get any polarization we wanted. If the
polarization of the down-converted photons is known, for example,
the crystal is cut for Tpye-1 SPDC, there is no need to use PBS.

\begin{figure}[b]
\includegraphics[width=8cm]{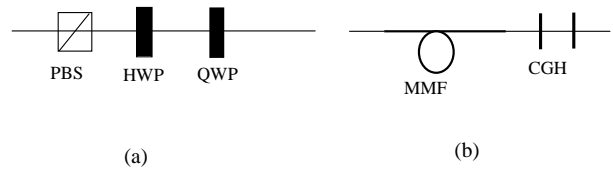}
\caption{(a) is a PC which is used to control the polarization of photons.
It includes PBS, HWP and QWP. (b) is an OAMC which is used to control the
OAM of Photons. It includes MMF and CGHs. The PBS (MMF) can be taken out if
we know the polarization (OAM) of the input photon.}
\end{figure}

With PCs, we can change the polarization of the photons according
to the paths they taken. For instance, photons with polarization
showing in Tab. 1 can be achieved. In this case, the state of Eq.
(2) will change to:

\begin{equation}
\left| \psi \right\rangle =\frac 1{\sqrt{3}}(\left| 0_H0_H\right\rangle
+\left| 1_H1_V\right\rangle +\left| 2_V2_V\right\rangle) .
\end{equation}

\begin{table}[tbp]
\caption{The polarization of photons in each arm of signal and idler path.}
\label{tabone}
\[
\begin{tabular}{|l|l|l|l|}
\hline
Signal path & Polarization & Idler path & Polarization \\ \hline
Short arm & H & Short arm & H \\ \hline
Medium arm & H & Medium arm & V \\ \hline
Long arm & V & Long arm & V \\ \hline
\end{tabular}
\]
\end{table}

A set of basis states can be defined with the polarization of the two photons\cite{Boganov04}%
:$\left| HH\right\rangle =\left| 0\right\rangle ;\left|
HV\right\rangle =\left| 1\right\rangle ;\left| VV\right\rangle
=\left| 2\right\rangle $. If just consider the information of
polarization, we get the state:
\begin{equation}
\left| \psi \right\rangle =\frac 1{\sqrt{3}}(\left| 0\right\rangle +\left|
1\right\rangle +\left| 2\right\rangle ).
\end{equation}
Obviously, we can control the polarization of photons in each path to change
this state to arbitrary qutrit: $\alpha \left| 0\right\rangle +\beta \left|
1\right\rangle +\gamma \left| 2\right\rangle $. If we additionally define $%
\left| VH\right\rangle =\left| 3\right\rangle $, we will get arbitrary
4-dimensional state: $\alpha \left| 0\right\rangle +\beta \left|
1\right\rangle +\gamma \left| 2\right\rangle +\delta \left| 3\right\rangle $.

Thus we can get the three-degree three-dimensional entangled GHZ
state:
\begin{equation}
\left| \psi \right\rangle =\frac 1{\sqrt{3}}(\left| 000\right\rangle +\left|
111\right\rangle +\left| 222\right\rangle ,
\end{equation}
where the first two terms of each $\left| nnn\right\rangle (n=0,1,2)$
represent the path information and the third term represents the
polarization information.

\subsection {Generation of entangled qutrits and four-degree three-dimensional GHZ state}
Now we will discuss the function of OAM control(OAMC). The
OAMC(see Fig. 2b) is a device that can be used to control the OAM
of photons. It has been shown that paraxial Laguerre-Gaussian(LG)
laser beams carry a well-defined orbital angular
momentum\cite{Allen92}, and the LG modes form a complete Hilbert
space. Every photon of the pure $LG_p^l$ mode beam carries an OAM
of $l\hbar $. Many experimental
works\cite{Mair01,Vaziri02,Vaziri03} about qudits are based on OAM
of the photons since the down converted photons from SPDC are
entangled in OAM. In these works, computer generated hologram(CGH)
and mono-mode fiber(MMF) were used to generate and detect high
order LG modes. Only the photons with $0$ OAM can passed the MMF,
all other modes have a large spatial extension, and therefore
cannot be coupled into MMF. The function of MMF to OAM is like PBS
to polarization. The CGHs are used to change the OAM of photons.
By shifting the CGHs\cite{VaziriJ} or designing special CGHs\cite
{Arlt98}, we can generate superposition of different OAM states.
The OAMC can be realized by the combination of MMF and CGHs. The
efficiency of this kind OAMC is lower than $1$.

By the help of OAMCs, and just consider the information of OAM of
the photons, we can engineer any pure two-partite
three-dimensional entangled states:

\begin{eqnarray}
\left| \psi \right\rangle &=&\left| 0\right\rangle (\alpha _{00}\left|
0\right\rangle +\alpha _{01}\left| 1\right\rangle +\alpha _{02}\left|
2\right\rangle )  \nonumber \\
&&+\left| 1\right\rangle (\alpha _{10}\left| 0\right\rangle +\alpha
_{11}\left| 1\right\rangle +\alpha _{12}\left| 2\right\rangle )  \nonumber \\
&&+\left| 2\right\rangle (\alpha _{20}\left| 0\right\rangle +\alpha
_{21}\left| 1\right\rangle +\alpha _{22}\left| 2\right\rangle ),
\end{eqnarray}
where $\left| 0\right\rangle ,\left| 1\right\rangle $ and $\left|
2\right\rangle $ is the OAM of the photons.

In detail, pure two-partite three-dimensional entangled states can
be divided into three group. First, it only includes the terms in
one of the three lines of Eq. (6), for example $\left|
0\right\rangle (\alpha _{00}\left| 0\right\rangle +\alpha
_{01}\left| 1\right\rangle +\alpha _{02}\left| 2\right\rangle )$.
We can get this type of states by controlling the photons taken
the signal path in the state $\left| 0\right\rangle $, and photons
taken the idler path in the state $\alpha _{00}\left|
0\right\rangle +\alpha _{01}\left| 1\right\rangle +\alpha
_{02}\left| 2\right\rangle $. Second, the state includes the terms
coming from two lines of Eq.(6), such as $\left| 0\right\rangle
(\alpha _{00}\left| 0\right\rangle +\alpha _{01}\left|
1\right\rangle +\alpha _{02}\left| 2\right\rangle )+\left|
1\right\rangle (\alpha _{10}\left| 0\right\rangle +\alpha
_{11}\left| 1\right\rangle +\alpha _{12}\left| 2\right\rangle )$.
This state can be generated when the photons taken the short arm
and medium arm of signal path have $0$ OAM, photons taken the long
arm of signal path have $1$ OAM, and photons taken the short arm
and medium arm of idler path are in the state $1/2(\alpha
_{00}\left| 0\right\rangle +\alpha _{01}\left| 1\right\rangle
+\alpha _{02}\left| 2\right\rangle )$, photons taken the long arm
of idler path is in the state $\alpha _{10}\left| 0\right\rangle
+\alpha _{11}\left| 1\right\rangle +\alpha _{12}\left|
2\right\rangle $. Third, the state includes terms coming from all
three lines of Eq. (6), for example $\left| 0\right\rangle (\alpha
_{00}\left| 0\right\rangle +\alpha _{01}\left| 1\right\rangle
+\alpha _{02}\left| 2\right\rangle )+\left| 1\right\rangle (\alpha
_{10}\left| 0\right\rangle +\alpha _{11}\left| 1\right\rangle
+\alpha _{12}\left| 2\right\rangle )+\left| 2\right\rangle (\alpha
_{20}\left| 0\right\rangle +\alpha _{21}\left| 1\right\rangle
+\alpha _{22}\left| 2\right\rangle ).$In this case, the photons
taken the signal short, medium and long arm of signal path can be
controlled to have $0,1,2$ OAM, and the photons taken the short,
medium and long arm of idler path in state $\alpha _{00}\left|
0\right\rangle +\alpha _{01}\left| 1\right\rangle +\alpha
_{02}\left| 2\right\rangle ,\alpha _{10}\left| 0\right\rangle
+\alpha _{11}\left| 1\right\rangle +\alpha _{12}\left|
2\right\rangle $ and $\alpha _{20}\left| 0\right\rangle +\alpha
_{21}\left| 1\right\rangle +\alpha _{22}\left| 2\right\rangle $
respectively. So we can get arbitrary pure two-partite
three-dimensional entangled state.

Consider the information of both the paths and OAM of photons, we
can obviously get the four-degree three-dimensional GHZ state:

\begin{equation}
\left| \psi \right\rangle =\frac 1{\sqrt{3}}(\left| 0000\right\rangle
+\left| 1111\right\rangle +\left| 2222\right\rangle ,
\end{equation}
where the first two terms of $\left| nnnn\right\rangle (n=0,1,2)$ carry the
information of the paths and last two terms carry the information of OAM of
photons.

\subsection {Generation of multi-dimensional entangled states}
It has been proven that angular momentum can be separate into
orbital and spin contributions\cite{Barnett02}. So we can use
OAMCs and PCs at the same time. In this situation, we can engineer
arbitrary pure three-degree three-dimensional entangled states:

\begin{equation}
\left| \psi \right\rangle =\sum_{m,n,p=0}^2\alpha _{mnp}\left|
mnp\right\rangle .
\end{equation}
Similarly,we can easily get the state:

\begin{equation}
\left| \psi \right\rangle =\frac 1{\sqrt{3}}(\left| 00000\right\rangle
+\left| 11111\right\rangle +\left| 22222\right\rangle .
\end{equation}
It is a five-degree three-dimensional maximally entangled GHZ state.

In fact, we will find these multi-degree multi-dimensional
entangled states can also be two-partite multi-dimensional
entangled states, while the dimension is $m\times n\times p\times
...$, where $m$ is the dimension of the first freedom, $n$ the
second, $p$ the third and so on. For example, we can do the
following definition: $\left| 0\right\rangle =\left|
0_H\right\rangle ,\left| 1\right\rangle =\left| 0_V\right\rangle
,\left| 2\right\rangle =\left| 1_H\right\rangle ,\left|
3\right\rangle =\left| 1_V\right\rangle ,\left| 4\right\rangle
=\left| 2_H\right\rangle ,\left| 5\right\rangle =\left|
2_V\right\rangle $, where $\left| 0_H\right\rangle $ represent the
OAM of the photon is $0$ and polarization $H$, others similarly.
This kind of states can be measured by combination of PBS, HWP,
QWP, CGH, and MMF. Then the state of Eq. (8) will change to a
two-partite six-dimensional($3\times 2$) entangled state:

\begin{equation}
\left| \psi \right\rangle =\sum_{m,n=0}^5\alpha _{mn}\left| mn\right\rangle .
\end{equation}
Obviously, if we can use more freedom of photons, we can easily
increase the dimension.

Bell inequalities and CHSH inequalities have been developed to
arbitrarily high-dimensional systems\cite{Gisin92,Collins02}. To
test these inequalities, we need to think the architecture to
increase the dimension of the systems. For example, in the OAM
schemes, lens must be used to get the maximally entangled states.
It will be difficult to realize if higher dimension is needed. The
present protocol shows that it seems much more convenient to
increase the dimension of systems using hyper-entanglement.
\section {conclusion}

In conclusion, different freedoms(path, polarization and OAM) of
down-converted photons from SPDC are used to generate
multi-dimensional entangled states. The photon pairs are entangled
in energy-time, polarization and OAM which is called hyper
entangled. We can engineer multi-dimensional states and
multi-dimensional entangled states of two partite systems in this
way. Multi-degree multi-dimensional GHZ states can also be
generated using this hyper entanglement. It is easier to control
this entangled qudits than we just use high dimension of single
freedom of photons. The hyper-entanglement is proved to be useful
for investigation of higher-dimensional entangled states.

\begin{center}
\textbf{Acknowledgments}
\end{center}

This work was funded by the Chinese National Fundamental Research Program
(2001CB309300), the National Natural Science Foundation(Grant No.60121503),
the NSF of China(10304017), the Innovation Funds from Chinese Academy of
Sciences.

\end{document}